\documentclass[sigconf]{acmart}

\AtBeginDocument{%
  }
\usepackage{tabularx}
\usepackage{graphicx} 
\usepackage{multirow}
\usepackage{enumitem}
\usepackage{makecell}

\usepackage{array}  
\usepackage{booktabs}
\usepackage{amsmath,amsfonts}
\usepackage{algorithm}
\usepackage{algpseudocode}
\usepackage{url} 
\usepackage{hyperref} 
\usepackage{textcomp}
\usepackage{xcolor}
\usepackage{siunitx} 
\usepackage{microtype}
\usepackage{soul}

\setcopyright{acmlicensed}
\copyrightyear{2018}
\acmYear{2018}
\acmDOI{XXXXXXX.XXXXXXX}
\acmConference[ ]{Make sure to enter the correct
  conference title from your rights confirmation email}{ }

\renewcommand\footnotetextcopyrightpermission[1]{} 
\setcopyright{none}                   
\begin{document}
\title{Beyond Task Completion: An Assessment Framework for Evaluating Agentic AI Systems}



\author{Sreemaee Akshathala}
\affiliation{%
  \institution{SERC, IIIT-Hyderabad}
  \city{Hyderabad}
  \country{India}}
\email{sreemaee.akshathala@research.iiit.ac.in}
\authornote{These authors contributed equally to this work.}

\author{Bassam Adnan}
\affiliation{%
  \institution{SERC, IIIT-Hyderabad}
  \city{Hyderabad}
  \country{India}}
\email{bassam.adnan@research.iiit.ac.in}
\authornotemark[1] 

\author{Mahisha Ramesh}
\affiliation{%
  \institution{SERC, IIIT-Hyderabad}
  \city{Hyderabad}
  \country{India}}
\email{mahisha.ramesh@research.iiit.ac.in}

\author{Karthik Vaidhyanathan}
\affiliation{%
  \institution{SERC, IIIT-Hyderabad}
  \city{Hyderabad}
  \country{India}}
\email{karthik.vaidhyanathan@iiit.ac.in}

\author{Basil Muhammed}
\affiliation{%
  \institution{MontyCloud Inc}
  \city{Bangalore}
  \country{India}}
\email{basil@montycloud.com}

\author{Kannan Parthasarathy}
\affiliation{%
  \institution{MontyCloud Inc}
  \city{Bangalore}
  \country{India}}
\email{kannan@montycloud.com}

\renewcommand{\shortauthors}{Akshathala, Adnan, et al.}

\begin{abstract}
Recent advances in agentic AI have shifted the focus from standalone Large Language Models (LLMs) to integrated systems that combine LLMs with tools, memory, and other agents to perform complex tasks. These multi-agent architectures enable coordinated reasoning, planning, and execution across diverse domains, allowing agents to collaboratively automate complex workflows. Despite these advances, evaluation and assessment of LLM agents and the multi-agent systems they constitute remain a fundamental challenge. Although various approaches have been proposed in the software engineering literature for evaluating conventional software components, existing methods for AI-based systems often overlook the non-deterministic nature of models. This non-determinism introduces behavioral uncertainty during execution, yet existing evaluations rely on binary task-completion metrics that fail to capture it. Evaluating agentic systems therefore requires examining additional dimensions, including the agent’s ability to invoke tools, ingest and retrieve memory, collaborate with other agents, and interact effectively with its environment. 
\noindent These challenges emerged during our ongoing industry collaboration with MontyCloud Inc.\footnote{https://montycloud.com/}, an industrial partner in the domain of Autonomous CloudOps, when we deployed an agentic system in production. These limitations surfaced during deployment, highlighting practical gaps in the current evaluation methods and the need for a systematic assessment of agent behavior beyond task outcomes. Informed by these observations and established definitions of agentic systems, we propose an end-to-end Agent Assessment Framework with four evaluation pillars encompassing LLMs, Memory, Tools, and Environment. We validate the framework on a representative Autonomous CloudOps use case, where experiments reveal behavioral deviations overlooked by conventional metrics, demonstrating its effectiveness in capturing runtime uncertainties.
\end{abstract}

\vspace{-10pt}

\keywords{Agentic AI, Assessment, Evaluation, AI-enabled systems, Software Engineering, CloudOps}

\maketitle
\begin{small}
\begin{table*}[!htbp]
\centering
\small
\caption{Uncertainties and Evaluation Metrics}
\label{tab:uncertainty_metrics_by_pillar}
\renewcommand{\arraystretch}{1.05}
\setlength{\tabcolsep}{6pt}
\begin{tabular}{|p{1.8cm}|p{1.8cm}|p{5.5cm}|p{6cm}|}
\hline
\textbf{Source} & \textbf{Uncertainty} & \textbf{Description} & \textbf{Evaluation Metrics} \\
\hline
\multirow{2}{=}{\textbf{LLM}} 
& Instruction\newline Following 
& Agent adherence to  policy constraints
& Instruction adherence score; Sequence correctness; Test case coverage; Judge scores \\
\cline{2-4}
& Safety \&\newline Alignment 
& Generated actions' compliance with safety policies and intended objectives
& Safety violation count; Policy compliance rate; Judge scores \\
\hline

\multirow{2}{=}{\textbf{Memory}} 
& Storage 
& Correctness and efficiency of memory updates and propagation
& Update correctness rate; Update latency; Duplicate entry count \\
\cline{2-4}
& Retrieval 
& Accuracy and completeness of retrieved information from memory systems
& Precision; Recall; F1-score; BLEU-1; ROUGE; Retrieval latency; Coverage ratio \\
\hline

\multirow{4}{=}{\textbf{Tools}} 
& Tool Selection 
& Correct identification of appropriate tool for given task
& Classification accuracy; Tool selection; Judge scores \\
\cline{2-4}
& Parameter \newline Mapping

& Semantic and contextual accuracy of tool parameters
& Parameter accuracy; Semantic correctness score; Judge scores \\
\cline{2-4}
& Tool \newline Sequencing 
& Correct execution order respecting dependencies and workflows
& Sequence correctness score; Phase completion ratio; Diagnostic-before-action adherence \\
\cline{2-4}
& Error \newline Interpretation 
& Agent understanding and response to tool execution errors
& Recovery success rate; Corrective action accuracy; Judge scores \\
\hline

\multirow{1}{=}{\textbf{Environment}} 
& Resource \newline Limitations
& Environment limitations like resource fluctuations, constraint changes and action authorization
& Guardrail violation count; Blocked action attempts; Authorization failures; Judge scores \\
\hline
    \end{tabular}
\end{table*}
\end{small}

\section{Introduction}
\label{sec:intro}
The evolution of LLMs over the last few years has shifted the AI system design from independent and isolated model deployment to agentic system architectures that integrate reasoning, planning, and tool use \cite{2024responsibleagents}. While multi-agent systems have a well-established foundation\cite{10.5555/1483085}, modern agentic systems extend these principles through memory mechanisms and environment interaction. This enables its application across diverse domains such as software engineering, customer service, and autonomous operations. This evolution requires the evaluation mechanisms to shift from model-centric metrics to system-level assessments. The integration of multiple components introduces significant evaluation challenges that conventional assessment methodologies do not address. The primary challenge in evaluating agentic systems stems from the inherent uncertainty \cite{10.1007/978-3-032-02138-0_14} introduced by non-deterministic AI models. Agentic systems exhibit variability in execution paths, tool selection, and memory retrieval patterns, causing uncertainty to propagate through the system, affecting the correctness of tool invocations, the consistency of memory operations, and the reliability of multi-agent interactions. 

\begin{figure}[!htbp]
\centering
\includegraphics[width=0.8\linewidth, height=0.5\textheight, keepaspectratio]{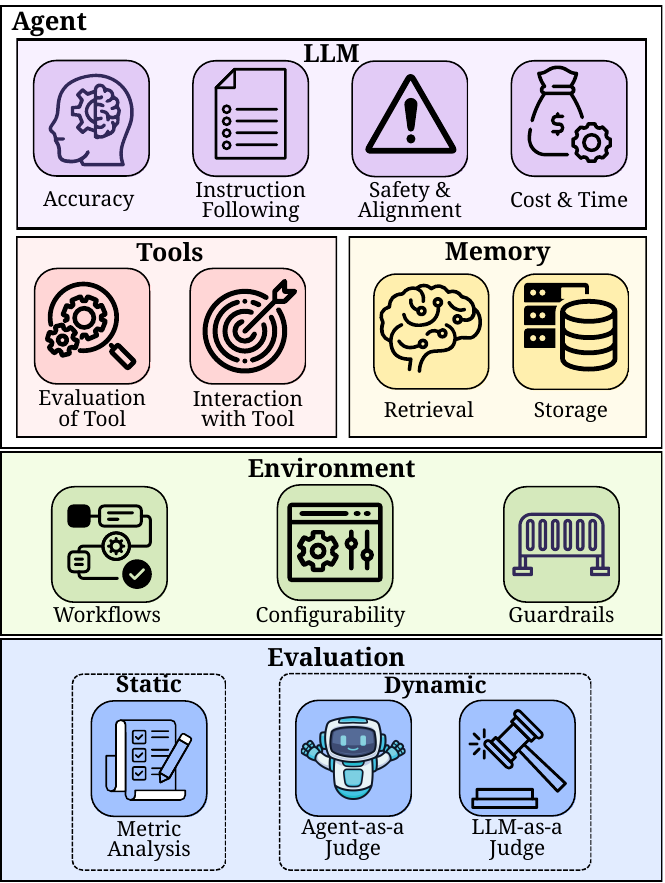}
\vspace{-10pt}
\Description{Overview of framework}
\caption{Agent Assessment Framework Overview}
\label{fig:main-overview}
\vspace{-10pt}
\end{figure}

\noindent Current research on agentic AI systems concentrates primarily on design patterns and architectural frameworks\cite{2025agenticroadmap}, with limited attention to systematic assessment methodologies. This gap became evident when we tested an agentic system in a CloudOps production environment at Montycloud. While the agents completed the tasks, deviations from expected policies or validation flows remained undetected by existing assessment methods as they surfaced only during runtime execution rather than in static validation phases.  
\noindent Based on these production-level insights, we designed a comprehensive assessment framework for evaluating agentic AI systems across the four identified pillars: LLM, Memory, Tools, and Environment. These pillars were derived based on the fundamental definition of an agent, an entity equipped with a reasoning component (LLM), memory, and tools for interaction with its environment\cite{2024agentopsobservability, moshkovich2025taminguncertaintyautomationobserving}. Based on existing research in self-adaptive systems \cite{weyns2018engineering, hezavehi2021uncertainty}, we also recognize the environment as a source of uncertainty which influences agent behavior. The framework integrates static, dynamic, and judge evaluation modes to capture behavioral failures beyond task success rates. Table~\ref{tab:uncertainty_metrics_by_pillar} provides an overview of the pillar-specific uncertainties and corresponding evaluation metrics that form the foundation of our assessment methodology.

\noindent We validate the assessment framework through experimental evaluation on MOYA, a multi-agent framework previously developed \cite{11030040}. With insights from Montycloud, we designed three production-motivated CloudOps scenarios of varying complexity, representing real operational challenges encountered in enterprise-level cloud management. Results demonstrate that while baseline metrics report task completion, our framework-specific assessments reveal substantial behavioral failures across all pillars, particularly in tool orchestration and memory retrieval.

\noindent The rest of the paper is organised as follows: Section \ref{sec:motivation} motivates the need for assessment through Autonomous CloudOps examples, Section \ref{methodology} introduces the framework architecture across all four pillars and the evaluation components. Section \ref{sec:exp} details the experimental setup and results, while sections \ref{sec:disc}, \ref{sec:rel}, and \ref{sec:conclusions} discuss the implications for researchers and practitioners, related work, and future work. Our experiments and results are available on GitHub\footnote{https://github.com/sa4s-serc/asf}. 

\section{Motivation}
\label{sec:motivation}
CloudOps manages and optimizes cloud infrastructure through automated workflows. Organizations navigating increasingly complex cloud environments require autonomous systems capable of executing diverse operational tasks at scale. CloudOps platforms like \textit{MontyCloud Inc.}\footnote{https://montycloud.com/} leverage AI agents to automate cloud management tasks, including resource provisioning, compliance enforcement, and incident response \cite{11030040}. These systems demonstrate the growing reliance on autonomous decision-making in operational workflows. Conventional evaluation frameworks largely focus on task outcomes and operational efficiency, overlooking behavioral failures which emerge at runtime. Agents might appear to perform well while violating policies, making uninformed decisions, or skipping verification checks. Evaluating behavioral reliability requires verifying whether agents consult policies before acting, perform diagnostic checks, retrieve relevant context, invoke tools with correct parameters, and respect guardrails that prevent unauthorized operations.

\noindent Due to the nature of agentic systems, the uncertainties extend beyond the performance of the model and span multiple sources, as shown in Table~\ref{tab:uncertainty_metrics_by_pillar}. These pillars capture an agent’s reasoning, memory, tool use \cite{2024agentopsobservability}, and include the environment as a source of operational uncertainty \cite{weyns2018engineering, hezavehi2021uncertainty}. The \textit{LLM pillar} introduces failures in instruction following and safety alignment. The \textit{Memory pillar} presents challenges in storage consistency and retrieval accuracy. The \textit{Tools pillar} exhibits failures in Tool selection, parameter mapping, and execution sequencing. The \textit{Environment pillar} represents the operational context in which agents are assessed, defined through structured workflows, guardrails for safe operation, and configurability for reproducible evaluation, assuming sufficient observability for monitoring and analysis. These limitations motivate a multi-layered evaluation strategy that examines agent behavior across complementary dimensions, including static, dynamic, and qualitative analyses, to enable assessment of behavioral reliability. The following subsections detail the framework components and evaluation strategies for each pillar.


\section{Agent Assessment Framework}
\label{methodology}

\begin{figure*}[!htbp]
\centering
\includegraphics[width=0.8\textwidth, keepaspectratio]{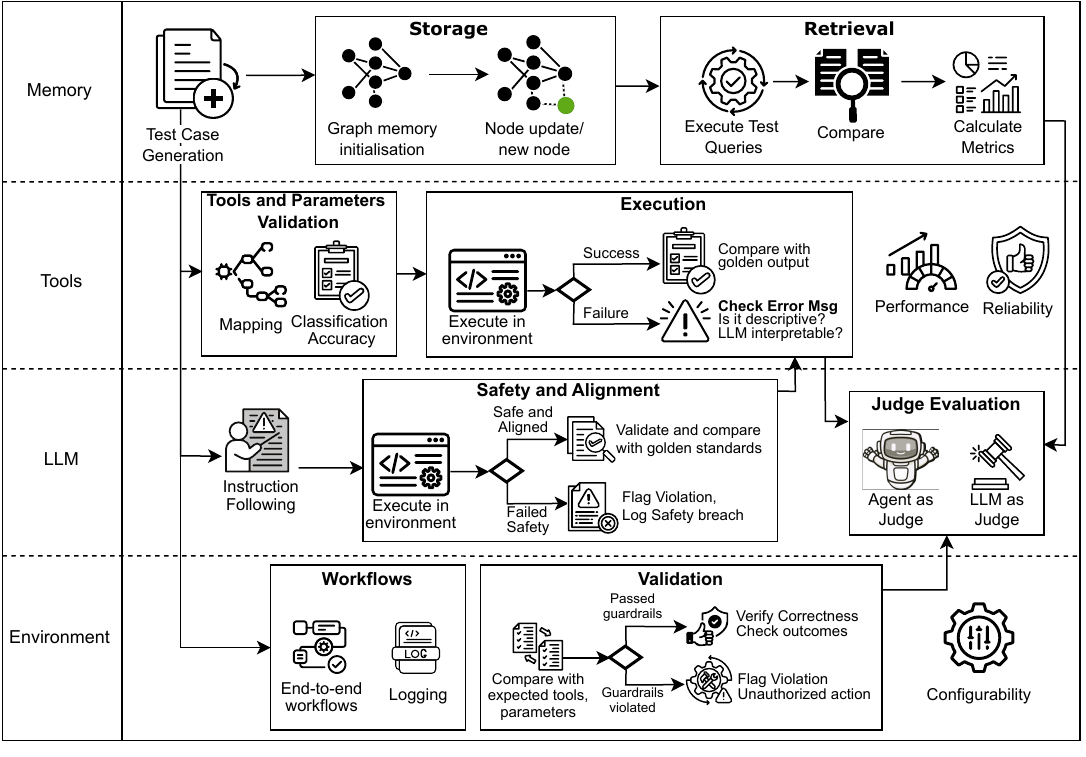}
\vspace{-15pt}
\Description{detailed Overview of framework}
\caption{Detailed view of Agent Assessment Framework}
\label{fig:approac}
\end{figure*}

\noindent  Figures \ref{fig:main-overview} and ~\ref{fig:approac} illustrate the proposed framework for evaluating agentic AI systems across four pillars, namely, \textit{LLM, Memory, Tools, and Environment}. The framework performs three layers of analysis: (i) \textit{Static Analysis}, which validates agent behaviors against predefined ground-truth specifications; (ii) \textit{Dynamic Execution}, which monitors runtime behaviors to detect deviations and policy violations; and (iii) \textit{Judge-based Evaluation}, which qualitatively assesses reasoning, safety, and alignment through LLM or Agent-as-a-Judge protocols. Together, these layers establish a comprehensive process for assessing both deterministic correctness and behavioral reliability in agentic systems.

\noindent The framework follows a modular structure where each pillar defines its own evaluation metrics and assessment focus. Results from the static and dynamic assessments are combined to measure behavioral correctness and reliability, while qualitative evaluations are used to assess reasoning and alignment. This structure allows the framework to provide both granular and system-level insights into agent performance across diverse operational contexts.\\
\noindent \textbf{Test Case Generation:} To ensure consistency across all pillars, the framework includes a \textit{test case generator} component that defines controlled scenarios for evaluation. Each test case specifies context, expected behaviors, and success criteria derived from ground-truth specifications. These cases are tailored to each pillar's objectives. For example, for the \textit{LLM} pillar, they test instruction following and safety alignment; for \textit{Memory}, retrieval accuracy and context retention; for \textit{Tools}, proper selection and parameter mapping; and for the \textit{Environment}, workflow execution and guardrail compliance. The generated test cases are used throughout static validation, dynamic monitoring, and judge-based evaluation to ensure systematic and reproducible assessment across all dimensions of agent performance.

\subsection{Memory}
A fundamental limitation of the agent's backbone LLM is its context length and its effective utilization. This is mitigated by providing the agent with a memory mechanism \cite{2024memorysurvey} that allows the agent to store and retrieve memories. Current state-of-the art approaches rely on extracting relations between entities and store them in a graph or vector-based storage engines \cite{mem02025}. Our Static analysis for the pillar computes precision, recall, F1-score, and BLEU-1 against gold-standard labels, followed by Judge evaluation.

\noindent \textbf{Storage:} Evaluates how effectively the agent structures, updates, and manages stored information such as events, conversations, and contextual data to ensure efficient and accurate memory retrieval. This evaluation focuses on whether these memory operations are performed correctly and efficiently, ensuring that obsolete information is replaced rather than duplicated.  Additionally, this component measures update latency, i.e., how long it takes for new information to propagate through the memory hierarchy and become visible to all agents. For example, in CloudOps practice, this may be observed when outdated configuration details of the same resource are retained, leading to multiple conflicting entries for the same resource.


\noindent \textbf{Retrieval:} Evaluates how accurately and efficiently the agent recalls relevant information from memory systems such as graph or vector-based stores, conversation history, or external knowledge sources. This component measures recall accuracy, which reflects how well the retrieved information aligns with the intended query. The evaluation spans four key retrieval types---Single-hop, Multi-hop, Temporal Reasoning, and Open-Domain. Single-hop tests the recall of facts from a single point of context; Multi-hop assesses retrieval across multiple interconnected memory instances; Temporal Reasoning examines the agent’s chronological understanding of events; and Open-Domain retrieval evaluates how effectively the agent combines conversational memory with external knowledge bases such as vector databases or cloud documentation. Effective retrieval ensures that the agent can synthesize context from multiple sources to provide consistent, contextually grounded responses. In CloudOps, this may involve retrieving recent system configuration updates and aligning WAFR\footnote{https://techreformers.com/solutions/wafr/} workload details with cloud best-practice guidelines to support informed operational decisions.


\subsection{Tools}

Tools enable agents to act within real or simulated environments, defining their operational competence. Within an agentic framework, they may be internal functions or external components defined through protocols like the Model Context Protocol (MCP) \cite{2025mcptoolbench}. Evaluating both the tool quality and the agent’s ability to use tools correctly is therefore essential, as failures often stem from ambiguous descriptions, incorrect parameter mappings, or inadequate error messages.\\
\noindent \textit{Quantitative Validation:} Based on the trajectory of an agent, the tool usage can be extracted and compared to golden labels as part of the Static analysis. This can be categorized into:\\
\noindent \textbf{Tool Classification Accuracy:} Measures if the correct tool is identified for a given task based on the problem description and available tool catalog. A misclassification or wrong sequencing may lead to invoking an incorrect or unrelated tool, compromising task reliability. In a CloudOps setting, the agent might select the Audit tool instead of the Monitoring tool due to incorrect classification.\\
\noindent \textbf{Parameter Accuracy:} Verifies that the parameters passed to a tool are semantically and contextually accurate. For instance, in CloudOps, an agent may attempt to scale a compute instance using a region name instead of an instance ID, leading to incorrect API calls or failed actions.\\
\noindent \textit{Qualitative Validation:} To assess how tools behave during real or simulated execution and how effectively the agent responds to runtime conditions. Tools are executed within a sandboxed environment that replicates operational conditions to verify whether they complete successfully.\\
\noindent \textbf{Error Message Evaluation:} When a tool call fails, the returned error should be interpretable and descriptive enough for the LLM to understand the cause and guide corrective actions. For instance, instead of having "Operation Failed" when the agent fails to delete an EC2 instance due invalid authorization, the error message should be descriptive, for example: "Insufficient IAM permissions for deleting EC2 instance". This in turn  reduces repetitive, invalid attempts.\\
\noindent \textbf{Performance:} Compares functionally equivalent tools or endpoints based on latency, cost, and consistency. In CloudOps, an agent may choose between invoking an API through a US-West or US-East server. Evaluating both paths under varying workloads helps determine the optimal route for cost-performance trade-offs.\\
\noindent \textbf{Reliability:}  Tracks error frequency and recovery behavior to assess whether tools execute consistently and recover gracefully from partial failures. Modular tools tend to be more reliable since they isolate faults, support independent validation and fault recovery. In contrast, monolithic tool definitions combining multiple operations introduce multiple points of failure\cite{2024shortcuts}. In CloudOps, this can manifest as API calls intermittently failing or returning inconsistent responses.

\subsection{LLM}
An LLM is a core part of the agent, and in extension, the framework as well. It is well integrated in various layers of the system which includes classification of tools, segregation of tasks from user query, LLM as a judge for evaluation, handling errors etc. This requires this component to be tested across various aspects. Our assessment's Static analysis verifies the presence of required keywords in memory queries prior to action execution. As depicted in Fig. \ref{fig:approac}, the rest of the assessment of the LLM pillar can be broadly categorized into the following components: \\
\noindent \textbf{Instruction Following}: Evaluates how accurately the LLM adheres to predefined instructions and rule-based workflows. This includes compliance with the system prompt and task-specific instructions, such as parsing execution logs, analyzing tool invocation sequences and memory queries, and verifying alignment with the expected flow and policy constraints. The evaluation is  performed using the proposed metrics - through static comparison of expected versus actual behavior, LLM-as-a-judge assessments of logs to detect attempted actions, and dynamic test case generation designed to measure the agent’s ability to follow instructions under varying conditions. In the CloudOps context, this may occur when an agent attempts to delete an EC2 instance despite restricted permissions, or bypasses mandatory steps in a workflow.\\
\noindent \textbf{Safety and Alignment:} Evaluates whether the agent’s generated actions or code are safe, compliant, and aligned with intended objectives. This is especially crucial when agents can execute or modify code and interact with live environments. The evaluation framework operates in two phases: static analysis, to detect potential risks before execution, and dynamic analysis, involving Judge Evaluation to assess runtime behavior. This includes validating outputs against safety policies, flagging violations, and logging breaches for traceability. In the CloudOps context, this may involve scenarios where the agent’s generated deployment commands contain unsafe operations such as modifying production configurations or bypassing approval checks.

\noindent \textbf{LLM-as-Judge} LLM-as-Judge \cite{2025llmjudge} evaluates task completion, safety, reasoning quality, memory usage, and policy correctness using structured prompts. The judge receives execution logs and produces scored assessments with justifications, capturing qualitative aspects beyond deterministic metrics.

\noindent \textbf{Agent-as-Judge}
Recent advancement in agent communication protocols\cite{2025a2a} enables  agent capability discovery via the \textit{Agent Card}\footnote{https://developers.googleblog.com/en/a2a-a-new-era-of-agent-interoperability/} which can be provided to an auditor agent that designs capability tests from the card. The auditor identifies capabilities, issues test instructions, and verifies execution through environment inspection tools that check tool logs, memory queries, and state changes \cite{2024agentsjudge}.
\subsection{Environment}
The environment in which an agent operates can be either simulated or real-world. Although real-world environments may offer more representative conditions, their variability often reduces reproducibility. To perform the following evaluations, the framework assumes that the environment provides  observability mechanisms to enable effective monitoring and analysis of the agent’s internal states, decisions, and interactions during execution. Modules such as telemetry, tracing, logging, and control-flow visibility collectively support continuous monitoring and facilitate root cause analysis (RCA) during fault conditions. Telemetry modules \cite{2024agentopsobservability} evaluate agents at runtime by capturing traces and metadata across multiple spans throughout the execution lifecycle, thereby improving explainability by making the system’s decision-making process interpretable. Monitoring helps identify changes within the environment during workflow execution, especially in CloudOps settings, where fluctuations can significantly impact operational costs. Strands Agents\footnote{ https://strandsagents.com/latest/documentation/docs/user-guide/observability-evaluation/observability/} apply observability principles at the agent level, whereas cloud platforms such as \textit{CloudTrail} provide system-wide monitoring and operational auditing.
The agent’s behavior is then evaluated within this structured setup, which enables controlled execution, supports repeatable experimentation, and allows verification of agent fidelity under different operational conditions. The desired characteristics of this environment include: \\
\noindent \textbf{Workflows:}
The environment contains predefined workflows or test cases that enable systematic assessment \cite{2025are}. These workflows can be defined in consultation with domain experts or system stakeholders to ensure they represent realistic operational scenarios and policy constraints. They can vary in complexity from simple single-tool invocations to multi-step chains involving interdependent tools and varied solution paths. The evaluation examines whether the agent preserves intended execution order, passes intermediate results correctly, and handles failures or unexpected responses properly throughout the workflow. In CloudOps, test cases might include scenarios of EC2 deployment with networking configuration or S3 bucket setup with access policies, where the evaluation verifies the agent selects the correct AWS APIs and provides accurate parameters in the expected order.

\noindent \textbf{Configurability:}
The environment provides mechanisms to configure, manage, and reset evaluation settings with minimal effort. When running multiple test cases, it should allow resetting to a known state without requiring a complete restart. Users can define upper and lower bounds for parameters such as resource allocation and cost limits, and evaluate the agent’s performance under varied workloads and fault injection scenarios \cite{2025aiops, 2025are}. This capability enables fine-grained experimentation and reproducibility, allowing the agent to be tested under diverse constraints and configurations. In CloudOps, this involves resetting environments to a baseline or intermediate state between tests, injecting faults, and setting cost or resource limits to enable systematic evaluation without manual cleanup in the console.

\noindent \textbf{Guardrails and Security:}
The environment enforces operational boundaries and access controls that define the limits within which the agent operates. These ensure that the agents or tools perform only permitted actions and prevent it from executing operations that could cause unintended or harmful outcomes. This involves the enforcement of predefined permissions and policies that govern the agent’s behavior and interactions within the system. In CloudOps, an agent’s attempt to invoke a tool for deleting an existing EC2 instance must be disallowed unless explicit authorization or approval is granted, typically enforced through IAM policies. Such restrictions should be deterministically implemented, either through logical constraints within the agent or system-level mechanisms like role-based access control. Defining these guardrails enables systematic detection and analysis of agent rule or policy violations, which can be monitored using open-source solutions such as Arize Guards \cite{2024agentopsobservability}.

\section{Experimentation and Results}
\label{sec:exp}

We evaluate our assessment framework through experimentation on an autonomous CloudOps system built using MOYA \cite{11030040}. Building on our motivation in Section~\ref{sec:motivation},  we addresses two primary objectives: effectiveness in capturing assessment dimensions across all four pillars, and efficiency in terms of cost and time. We investigate the following research questions:

\begin{itemize}
    \item \textbf{RQ1: How effective is the proposed framework in capturing agent assessment dimensions compared to baseline evaluations?} We compare our four-pillar framework against a baseline evaluation that captures only task completion and tool usage ratios, measuring the framework's ability to surface failures overlooked by conventional approaches.
    \item \textbf{RQ2: What are the failure patterns across assessment pillars?} We analyze the distribution and frequency of failures across the pillars to characterize the types of agent failures.
    \item \textbf{RQ3: What is the efficiency of the framework in terms of cost and time?} We quantify the computational overhead through scenario execution costs, token consumption, and evaluation time for both runtime execution and judge-based assessment protocols.
\end{itemize}

\subsection{Experimental Setup}
For the experimental setup, we extended MOYA to incorporate memory mechanisms using the default settings of Mem0 \cite{mem02025} as the Memory pillar. Based on prior findings on the rule-following capabilities of different LLMs \cite{kcif2025, bif2024}, we selected the GPT-4o model for our experiments. with a temperature of 0.7, chosen after several trial runs to balance predictability and creativity in agent responses. We use ChromaDB for storing embeddings generated via text-embedding-3-small model. Based on production issues we encountered, we generated synthetic data consisting of instances, logs, and policies using Sonnet 4.5 due to its ranking on LMArena\footnotemark{}. These were exposed to the agents as callable tools generated, enabling them to retrieve and process information about simulated cloud resources and operational rules. We defined three workflows of varying complexity representing real-world CloudOps scenarios:

\begin{enumerate}
\item \textbf{Cost Optimization (S1):} A Cost Optimization Agent must reduce AWS infrastructure costs by 30\% by identifying and terminating underutilized EC2 instances while adhering to organization policies on non-production prioritization, environment verification, and CAB approval. Tests whether the agent balances cost reduction goals with operational safety constraints.

\item \textbf{Security Incident Response (S2):} A Security Remediation Agent must remediate publicly accessible S3 bucket containing sensitive data exposed for 5 days. The agent must assess exposure scope, check compliance requirements, enable logging before policy changes, and apply remediation while maintaining authorized service access. 

\item \textbf{Multi-Agent Root Cause Analysis (RCA) (S3):} Diagnose 60\% response time degradation in payment service caused by network misconfiguration 5 minutes prior. Requires coordination between Performance, Security, and RCA agents to identify root cause through temporal correlation. Tests whether the system identifies configuration issues versus treating symptoms through costly scaling.
\end{enumerate}
Each scenario was executed three times independently, with results averaged to ensure reliability and account for non-deterministic LLM behavior. We use API keys to run these experiments and report the token consumption and response time.

\footnotetext{https://lmarena.ai/leaderboard}

\subsection{Results}

We evaluate our assessment framework through three research questions. Table~\ref{tab:rq-metrics-overview} defines the metrics used across all evaluations, the trajectory and baseline evaluations are available on GitHub\footnote{https://github.com/sa4s-serc/asf}. We report the results for each scenario across the agents used.

\begin{table}[!htbp]
\setlength{\tabcolsep}{3pt}
\renewcommand{\arraystretch}{1.1}
\vspace{-10pt}
\caption{ Overview of the metrics 
computed through deterministic log analysis: tool sequences matched against 
phase-specific requirements, memory retrieval compared to golden labels 
(P/R/F1), and LLM adherence verified through policy lookup patterns in 
execution traces.}
\label{tab:rq-metrics-overview}
\centering
\small
\begin{tabular}{l|p{2cm}|>{\raggedright\arraybackslash}p{4.3cm}}
\toprule
\textbf{Category} & \textbf{Metric} & \textbf{Definition} \\
\midrule
\multirow{2}{*}{LLM} & Policy Adherence & Look up policy guidance before acting. \\
\cline{2-3}
 & Dependency Inquiry & Retrieving dependency context \newline before acting. \\
\hline
\multirow{3}{*}{Tools} & Tool usage & Ratio of tools invoked vs. required \newline tool list. \\
\cline{2-3}
 & Tool sequence & Runs respecting diagnostic-before-action order. \\
\cline{2-3}
 & Expected tool calls & Mandatory tools invoked with \newline required parameters. \\
\hline
Memory & Memory coverage & Required memory lookups issued \newline (Precision, Recall, F1). \\
\hline
\multirow{2}{*}{Environment} & Task completion & Runs meeting declared objective. \\
\cline{2-3}
 & Production guardrail & Flags for environment protection \newline violations. \\
\bottomrule
\end{tabular}
\end{table}

\subsubsection{\textbf{RQ1: How effective is the proposed framework in capturing agent assessment dimensions compared to baseline evaluation?}}

We compare our four-pillar framework against baseline evaluation, which captures only task completion and tool usage ratio as standard metrics in agent evaluation workflows verifying success conditions and token consumption. Our framework extends this through deterministic pillar-specific metrics (Table~\ref{tab:rq1-combined}) and two judge-based evaluation protocols: Agent-as-Judge and LLM-as-Judge.

For Agent-as-Judge evaluation, an auditor agent dynamically designs capability tests based on each worker agent's card, instructs the worker to demonstrate specified capabilities, then verifies execution through environment checks. Table~\ref{tab:agent-judge} reports the percentage of capability tests passed per agent. LLM-as-Judge provides holistic assessment across task completion, safety, and memory dimensions using structured prompts (Table~\ref{tab:llm-judge}).

\begin{table}[!htbp]
\vspace{-10pt}
\setlength{\tabcolsep}{2pt}
\renewcommand{\arraystretch}{1.1}
\caption{Baseline (B) vs. Framework (F) metrics across all scenarios (S1–S3)}
\vspace{-10pt}
\label{tab:rq1-combined}
\centering
\small
\begin{tabularx}{\linewidth}{l *{6}{>{\centering\arraybackslash}X}}
\toprule
& \multicolumn{2}{c}{\textbf{S1}} & \multicolumn{2}{c}{\textbf{S2}} & \multicolumn{2}{c}{\textbf{S3}} \\
\cmidrule(lr){2-3}\cmidrule(lr){4-5}\cmidrule(lr){6-7}
\textbf{Metric} & \textbf{B} & \textbf{F} & \textbf{B} & \textbf{F} & \textbf{B} & \textbf{F} \\
\midrule
Task completion (\%)     & 0    & 0    & 100  & 100  & 33   & 33 \\
Tool usage (ratio)       & 0.80 & 0.80 & 0.97 & 0.97 & 0.62 & 0.62 \\
\midrule
Tool sequence (\%)       & --   & 100  & --   & 33   & --   & 53 \\
Expected calls (\%)      & --   & 100  & --   & 67   & --   & 62 \\
Policy Adherence (\%)    & --   & 33   & --   & 100  & --   & 0 \\
Dependency Inquiry (\%)  & --   & 100  & --   & 0    & --   & 67 \\
\midrule
Memory precision (P) (\%)    & --   & 33.7 & --   & 30.0 & --   & 75.56 \\
Memory recall (R) (\%)       & --   & 37.9 & --   & 13.1 & --   & 50.67 \\
Memory F1 (\%)           & --   & 31.1 & --   & 18.1 & --   & 51.32 \\
\bottomrule
\end{tabularx}
\end{table}

Baseline evaluation reports identical metrics across both approaches (e.g., S2: 100\% task completion, 0.97 tool usage ratio). However, framework-specific metrics reveal underlying execution issues: S1 achieved perfect tool sequencing (100\%) but only 33\% policy adherence, indicating actions proceeded without consulting safety guidelines. Memory performance varied significantly, with S2 showing particularly low recall (13.1\%) and F1 score (18.1\%), while S3 achieved 51.32\% memory F1. Tool sequence correctness ranged from 33\% (S2) to 100\% (S1).

\begin{table}[!htbp]
\setlength{\tabcolsep}{2pt}
\renewcommand{\arraystretch}{1.1}
\caption{Memory mechanisms (for S3):}
\label{tab:rq1-multi-memory}
\centering
\small
\vspace{-10pt}
\begin{tabularx}{\linewidth}{l|c|c|X}
\hline
\textbf{Mechanism} & \textbf{P} & \textbf{R} & \textbf{Expected (gold)} \\
\hline
Single-hop & 63.3 & 61.1 &
Retrieve one key fact such as baseline response times, subnet, or Security Group (SG) configuration. \\
\hline
Multi-hop & 100.0 & 26.5 &
Link \emph{change} and \emph{topology} to confirm DB access from app subnet on port 5432. \\
\hline
Temporal & 100.0 & 29.8 &
Correlate symptom onset with SG change timing (within 5–10 minutes) to infer causality. \\
\hline
\end{tabularx}
\end{table}

Table~\ref{tab:rq1-multi-memory} decomposes memory retrieval performance across four mechanisms for S3. Single-hop retrieval achieved balanced precision (63.3\%) and recall (61.1\%), while multi-hop and temporal reasoning showed perfect precision (100\%) but substantially lower recall (26.5\% and 29.8\% respectively), indicating the agent retrieved accurate information when querying but failed to retrieve all relevant memories.

\begin{table}[!htbp]
\setlength{\tabcolsep}{4pt}
\renewcommand{\arraystretch}{1.1}
\vspace{-10pt}
\caption{LLM-as-Judge results across scenarios (\%)}
\label{tab:llm-judge}
\centering
\small
\vspace{-10pt}
\begin{tabular}{lccccc}
\toprule
\textbf{Scenario} & \textbf{Completion} & \textbf{Safety} & \textbf{Memory} & \textbf{Reasoning} & \textbf{Overall} \\
\midrule
S1 & 60 & 85 & 75 & 70 & 72 \\
S2 & 90 & 85 & 70 & 80 & 85 \\
S3 & 100 & 100 & 100 & 100 & 100 \\
\bottomrule
\end{tabular}
\end{table}

LLM-as-Judge assessment (Table~\ref{tab:llm-judge}) assigns overall scores of 72\% (S1), 85\% (S2), and 100\% (S3). Completion scores increased from 60\% (S1) to 100\% (S3), while safety scores remained consistently high (85-100\%). Memory usage scores ranged from 70\% (S2) to 100\% (S3), and reasoning scores improved from 70\% (S1) to 100\% (S3). The completion score in S1 is reported as 60\% since the agent expected user input before proceeding to take an action, which is why its completion rate is reported as 0\% in Table~\ref{tab:rq1-combined}.
\begin{table}[!htbp]
\vspace{-10pt}
\setlength{\tabcolsep}{3pt}
\renewcommand{\arraystretch}{1.1}
\caption{Agent-as-Judge evaluation}
\vspace{-10pt}
\label{tab:agent-judge}
\centering
\small
\begin{tabularx}{\linewidth}{l|c|X}
\hline
\textbf{Agent (Scenario)} & \textbf{Tests} & \textbf{Key Findings} \\
\hline
Cost Agent (S1) & 6/6 & Validated instances, adhered to policies. \\
\hline
Security Agent (S2) & 6/6 & Completed access control and compliance checks. \\
\hline
Performance Agent (S3) & 9/10 & Missed configuration lookup; correctly compared metrics and reviewed logs. \\
\hline
Remediation Agent (S3) & 4/5 & Skipped pre-policy check; maintained safe rollback. \\
\hline
RCA Agent (S3) & 6/6 & Coordinated workflow and identified root cause. \\
\hline
\end{tabularx}
\vspace{-10pt}
\end{table}
Agent-as-Judge evaluation (Table~\ref{tab:agent-judge}) shows test results across five agents. Cost Agent (S1) and Security Agent (S2) passed all capability tests, while S3's multi-agent scenario revealed failures.

\subsubsection{\textbf{RQ2: What are the failure patterns across assessment pillars?}}

We analyze failure patterns by tracking average failures per pillar across scenarios to identify which assessment dimensions would be missed without pillar-specific evaluation. Table~\ref{tab:rq2-ablation} reports the average number of failures observed in each pillar, with example failure modes illustrating the types of issues captured.

\begin{table}[!htbp]
\setlength{\tabcolsep}{3pt}
\renewcommand{\arraystretch}{1.1}
\caption{Pillar ablation: Average failures per pillar}
\label{tab:rq2-ablation}
\centering
\small
\vspace{-10pt}
\begin{tabular}{l|c|c|c|p{4.1cm}}
\hline
\textbf{Pillar} & \textbf{S1} & \textbf{S2} & \textbf{S3} & \textbf{Example Failure} \\
\hline
LLM & 1.67 & 1.33 & 1.00 &
Skipped policy validation before instance termination. \\
\hline
Tools & 1.00 & 2.33 & 7.67 &
Missed diagnostic or verification steps before applying remediation. \\
\hline
Memory & 0.67 & 2.33 & 3.67 &
Did not recall previous role mappings or configuration changes. \\
\hline
Environment & 0.00 & 0.00 & 2.00 &
Production instance state changed despite guardrail expectations.\\
\hline
\end{tabular}
\end{table}

The Tools pillar exhibited the highest failure rate in S3 (7.67 average failures), primarily due to missed diagnostic or verification steps before remediation actions. Memory failures increased with scenario complexity, from 0.67 (S1) to 3.67 (S3), reflecting difficulties in recalling previous role mappings and configuration changes. LLM pillar failures decreased from 1.67 (S1) to 1.00 (S3), with policy validation violations being the primary failure mode. Environment guardrail violations appeared only in S3 (2.00 average failures), where production instance states changed despite protection constraints.


\subsubsection{\textbf{RQ3: What is the efficiency of the framework in terms of cost and time?}}

\begin{figure}[!htbp]
\centering
\includegraphics[width=0.8\linewidth]{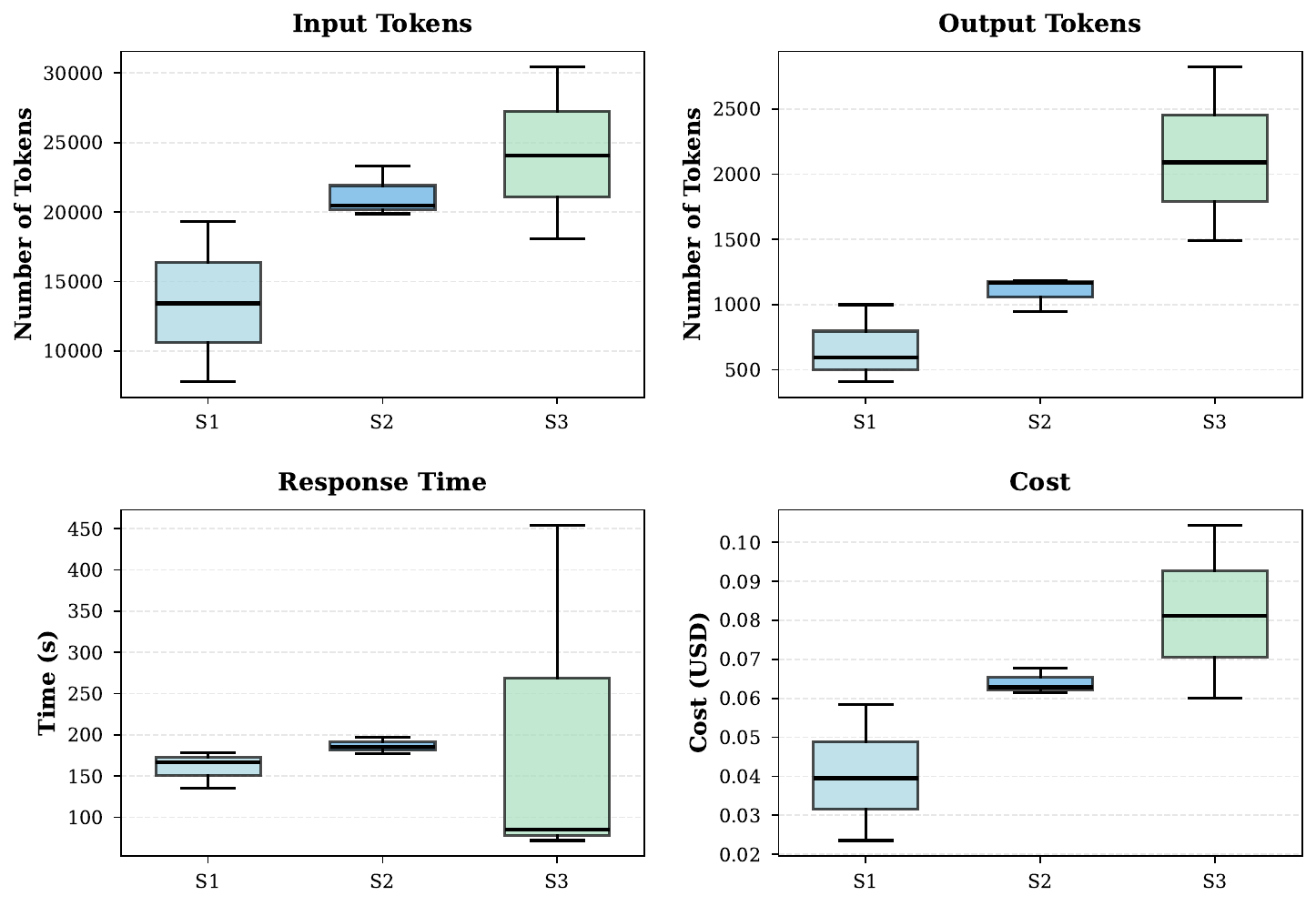}
\Description{efficiency analysis}
\caption{Distribution of input tokens, output tokens, response time, and cost across scenarios}
\label{fig:efficiency-analysis}
\vspace{-10pt}
\end{figure}

\begin{figure}[!htbp]
\centering
\includegraphics[width=\linewidth]{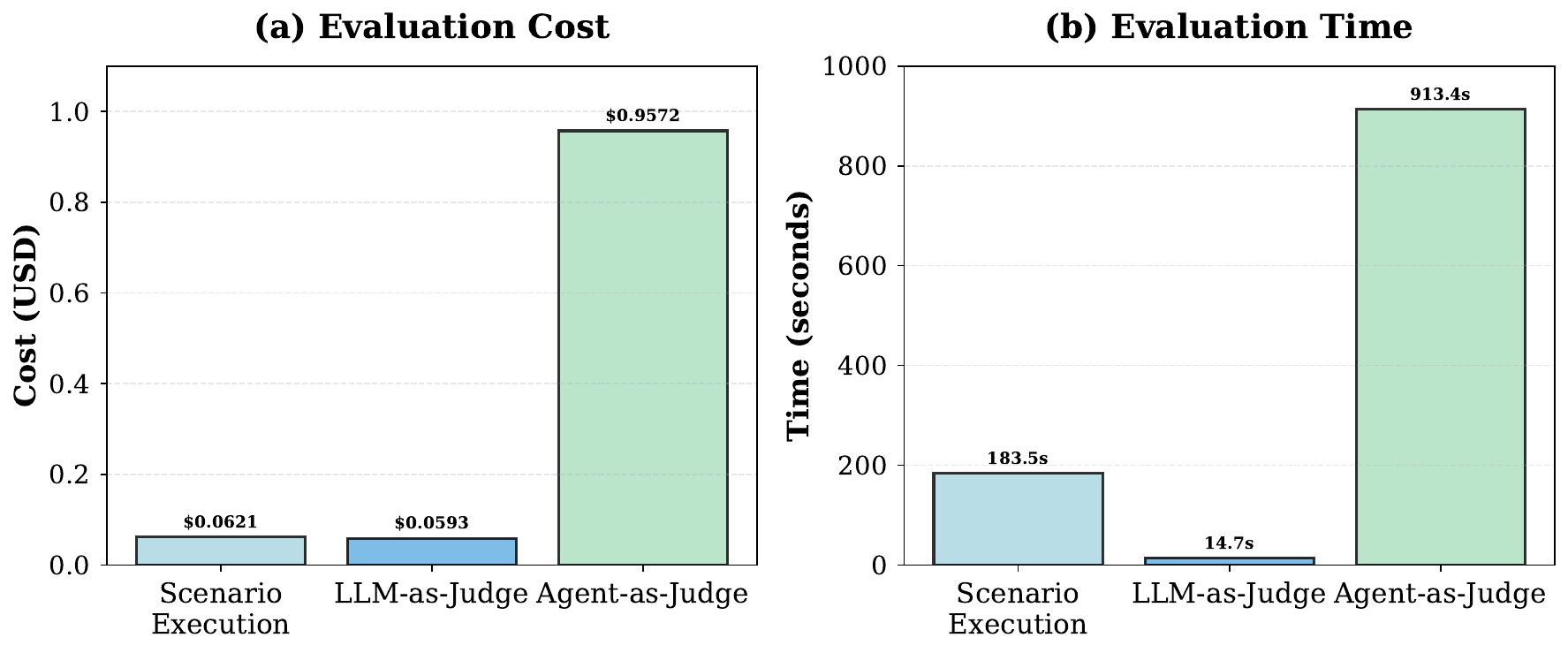}
\Description{A bar chart comparing the computational overhead in terms of cost and time for two different evaluation protocols: LLM-as-Judge and Agent-as-Judge. This alt-text explains the chart's content for accessibility.} 
\vspace{-20pt}
\caption{Evaluation overhead comparison showing (a) cost and (b) time for LLM-as-Judge and Agent-as-Judge protocols.}
\vspace{-10pt}
\label{fig:evaluation-costs}
\end{figure}

We measure the computational overhead introduced by our framework through scenario execution costs, token consumption, and evaluation time. Figure~\ref{fig:efficiency-analysis} shows the distribution of these metrics across three runs per scenario, while Figure~\ref{fig:evaluation-costs} compares the overhead introduced by LLM-as-Judge and Agent-as-Judge evaluation protocols.

\noindent Average scenario execution required \$0.0621 per run, consuming 19,644 input tokens and 1,301 output tokens over an average execution time of 183.5 seconds. Costs scaled with scenario complexity from S1 (\$0.0405, 160.3s) to S3 (\$0.0818, 203.5s). Fig. \ref{fig:efficiency-analysis} reveals variability in S1 input tokens (7,784--19,335) and S3 response time (72--454s), indicating scenario-dependent resource utilization patterns.

LLM-as-Judge added minimal overhead (\$0.0593 total across all scenarios, 14.7 seconds), consuming 18,602 input tokens and 1,277 output tokens. In contrast, Agent-as-Judge incurred higher costs (\$0.9572 total, 913.4 seconds) with 289,006 input tokens and 23,472 output tokens. The cost difference reflects the depth of assessment. S3 accounted for the majority of Agent-as-Judge cost (\$0.6110, 632.1s) due to multi-agent coordination requiring extensive verification cycles. Agent-as-Judge incurs 16× higher cost and 62× longer execution time due to extensive capability testing.

\section{Discussions}
\label{sec:disc}
Our collaboration with MontyCloud provided continuous feedback from production-scale deployments, allowing us to validate the practicality and efficiency of the proposed framework under real operational conditions. The comparison from the baseline (RQ1) shows that conventional approaches miss critical behavioral issues. For instance, S1 achieved perfect tool sequencing but only 33\% policy adherence, taking actions without consulting safety guidelines-a failure baseline metrics cannot detect. S2 showed low memory recall (13.1\%) despite task completion. Judge-based protocols revealed additional failures: the Remediation Agent skipped verification steps in 20\% of tests, while LLM-as-Judge gave S1 only 72\% overall. Further, in multi-agent scenario (S3) Judge methods help identify the error propagating through agents. This demonstrates that assessment frameworks must explicitly check instruction following, policy adherence, and memory usage beyond final outcomes. The pillar ablation study (RQ2) reveals where assessment effort should focus. Tool orchestration had the highest failure rate in complex scenarios, primarily from skipping diagnostic steps. Memory failures increased with scenario complexity, while Environment violations appeared only in multi-agent scenarios where production states changed despite guardrails. These patterns indicate that tool sequencing and memory management need particular attention in the multi-agent scenarios. The efficiency analysis (RQ3) shows practical feasibility with average execution cost of \$0.0621. LLM-as-Judge added minimal overhead (\$0.0593, 14.7s), while Agent-as-Judge incurred higher costs (\$0.9572, 913.4s). This suggests LLM-as-Judge for continuous monitoring and Agent-as-Judge for  pre-deployment audits.
\newline
\textbf{Implications for Researchers:} While existing approaches \cite{2025hasanempiricalstudytestingpractices} focus on model-centric metrics, there is a need for assessment of behavioral uncertainness across system components. Our results show the agents violating policies and missing important context across the scenarios evaluated in RQ1. This reveals research directions for handling the high tool orchestration failure rate in multi-agent scenarios indicates the need for better coordination mechanisms and formal verification methods that enforce diagnosis before taking an action. The precision-recall gap in memory presents the need for balancing accuracy with completeness, agents retrieve correct information but miss decision relevant context. The cost tradeoff discussed in RQ3 between the judge-based protocols motivates a research direction for dynamically adaptive evaluation strategies to select the assessment depth based on risk levels. Further, our scenarios required manual specification of expected behaviors while generating test cases, automated test case generation from agent capabilities presents a promising direction. Extending our framework to other agentic paradigms such as code generation or multi modal agents will validate and identify domain-specific evaluation requirements. 

\noindent\textbf{Implications for Practitioners:}
Existing methods evaluate primarily on final outcomes \cite{2025agenttrajectory} , which  which may explain the low translation from a prototype to production  conversion rates, due to which these failures surface which are otherwise invisible to standard metrics. Our results show that agents exhibit these failures, in S3 skipped policy checks, while S1 could not resolve the issue, it terminated while asking for permission to terminate the instances. This highlights the need for evaluations to perform behavior-based testing apart from outcome-based testing. Developers should implement pillar-specific checks for instruction following, policy adherence and implement defensive guardrails at the environment layer (IAM policies, control policies). For  evaluation, using LLM and Agent based judges while highlight such behavioral aspects at runtime.  Implanting telemetry by capturing tool call invocations, memory retrieval and storage, policy checks will set the foundation for prompt refinements and root cause analysis when agents fail.
\subsection{Threats to Validity}
\textbf{External Validity.} The evaluation focuses on CloudOps, limiting generalizability to other domains. The three scenarios may not capture challenges in creative tasks or long-horizon planning. To mitigate this to an extent we created these scenarios based on our experience on CloudOps and created the scenarios in varying complexity.

\noindent \textbf{Internal Validity.} Uniform temperature (0.7) and model selection (GPT-4o) may not capture full non-deterministic behavior distributions. Ground truth contracts defined by the us may not reflect all valid solution paths. Judge-based protocols introduce potential LLM evaluation biases.

\noindent \textbf{Construct Validity.} Metrics may not capture all reliability aspects. Memory assessment excludes storage efficiency and concurrent access evaluation. Environment checks do not measure observability quality or recovery mechanisms due to the textual simulation. 

\section{Related Work}
\label{sec:rel}
Research on agentic systems highlights the ongoing challenges across the domains of memory, task-following, safety, evaluation, and multi-agent coordination. In the memory domain, Maharana et al. \cite{maharana2024evaluating} show that LLMs struggle to track long conversations, while RAG-based approaches improve access to information there, Hu et al.  \cite{hu2025evaluating}  highlight its limitations. Overall, these findings suggest that a combination of approaches is needed to handle the evaluation of memory mechanism. For instruction-following, Qi et al. \cite{qi2025agentif} show the importance of assessment to capture  performance on complex instructions involving multiple constraints, especially when tools or conditions are involved.  Dong et al. \cite{2024agentopsobservability} and Andriushchenko et al. \cite{andriushchenko2024agentharm} focus on improving transparency, multi-agent coordination, and the robustness against harmful instructions. These studies emphasize the need for evaluation frameworks that balance reasoning, execution, safety, and monitoring. Further, research on self-adaptive and multi-agent systems by Nascimento et al. \cite{nascimento2023self} demonstrate the inconsistencies in memory and decision-making remain. The study “Why Do Multi-Agent LLM Systems Fail?” \cite{cemri2025multi} by Cemri et al. analyzes MAS failures by identifying gaps in specifications, alignment, and verification. Despite these advances, most evaluations focus on model accuracy or individual components, leaving runtime uncertainties, tool use, multi-agent coordination, and interactions with dynamic environments underexplored. To the best of our knowledge, there does not exist  unified evaluation system of an agentic system. To address these limitations, we propose an  Agent Assessment framework that evaluates agentic systems across four pillars: LLMs, Memory, Tools, and Environment, and highlight the failures which surface using   CloudOps scenarios to capture overall agent behavior.


\section{Conclusions and Future Work}
\label{sec:conclusions}


In this work, we present the first steps towards an assessment framework for evaluating agentic AI systems across four pillars. Through an industrial case of autonomous cloudOps, we demonstrate the effectiveness and efficiency of our framework. Future research directions include, but are not limited to: i) applying to diverse domains beyond CloudOps; ii) Integration with self-adaptive abilities to support continuous improvement of the agentic system; and iii) leveraging agent-as-judge to discover failure modes and assessment automation.

\bibliographystyle{acm}
\bibliography{abbrev ,references.bib}

\end{document}